\documentclass[a4paper,fleqn]{cas-sc}

\usepackage[authoryear]{natbib}
\usepackage[]{todonotes}
\usepackage{subfigure}
\usepackage{pdfpages}

\usepackage[nolist]{acronym}
\begin{acronym}[list of abbreviations]
\acro{adhd}[ADHD]{\textit{Attention Deficit Hyperactivity Disorder}}
\acro{aba}[ABA]{\textit{Applied Behavior Analysis}}
\acro{api}[API]{\textit{Application Programming Interface}}
\acro{asd}[ASD]{\textit{Autism Spectrum Disorder}}
\acro{ati}[ATI]{\textit{Affinity for Technology Interaction}}
\acro{bml}[BML]{\textit{Behavior-Markup-Language}}
\acro{dtt}[DTT]{\textit{Discret Trial Training}}
\acro{hlrc}[HLRC]{\textit{High-Level-Robot-Control}}
\acro{ip}[IP]{\textit{Internet Protocol}}
\acro{ja}[JA]{\textit{Joint-Attention}}
\acro{hmi}[HMI]{\textit{Human-Machine Interaction}}
\acro{hri}[HRI]{\textit{Human-Robot Interaction}}
\acro{rct}[RCT]{\textit{response-cost-token}}
\acro{ros}[ROS]{\textit{Robot-Operating-System}}
\acro{rdoc}[RDoC]{\textit{Research Domain Criteria}}
\acro{sdk}[SDK]{\textit{Standard-Development-Kit}}
\acro{sus}[SUS]{\textit{System Usability Scale}}
\acro{tcp}[TCP]{\textit{Transmission Control Protocol}}
\acro{tpb}[TPB]{\textit{Theory of Planned Behavior}}
\acro{vaco}[VACO]{\textit{Virtual Robotic Agent for Concentration}}
\acro{tts}[TTS]{\textit{Text-To-Speech}}
\acro{woz}[WoZ]{\textit{Wizard of Oz}}
\acro{xml}[XML]{\textit{eXtensible-Markup-Language}}
\acro{gui}[GUI]{\textit{Graphical-User-Interface}}
\end{acronym}
\def\tsc#1{\csdef{#1}{\textsc{\lowercase{#1}}\xspace}}
\tsc{WGM}
\tsc{QE}
\tsc{EP}
\tsc{PMS}
\tsc{BEC}
\tsc{DE}

\begin{document}
\let\WriteBookmarks\relax
\def\floatpagepagefraction{1}
\def\textpagefraction{.001}
\shorttitle{Virtual Robotic Agent for Concentration}
\shortauthors{B. Richter et~al.}

\title[mode = title]{VACO: a Multi-perspective Development of a Therapeutic and Motivational Virtual Robotic Agent for Concentration for children with ADHD}

\author[1]{Birte Richter}[
                        orcid=0000-0002-0957-2406]
\cormark[1]
\ead{birte.richter@uni-bielefeld.de, Bielefeld University, Universitatsstrasse 25, 33615 Bielefeld}
\credit{Conceptualization, Methodology, Software, Investigation, and  Formal analysis for Study 1-3; Writing - Original Draft, Visualization, Project administration} 
\address[1]{Bielefeld University, Medical School OWL,  Medical Assistance Systems}

\author[2]{Ira-Katharina Petras}
\credit{Conceptualization, Methodology, and Investigation for Study 1-3; Writing - Original Draft}
\address[2]{Bielefeld University, Medical School and University Medical Center OWL, Protestant Hospital of the Bethel
Foundation, Department of Child and Adolescent Psychiatry and Psychotherapy}
\author[3]{Anna-Lisa Vollmer}
\credit{Conceptualization, Methodology, and Investigation for Study 1; Writing - Original Draft}
\address[3]{Bielefeld University, Medical School OWL, Interactive Robotics in Medicine and Care}

\author[1]{Ayla Luong}
\credit{Investigation for Study 2-3;}

\author%
[2,4]{Michael Siniatchkin}
\credit{Conceptualization, Methodology, Writing - Review \& Editing , and Supervision}

\author%
[1,4]{Britta Wrede}
\credit{Conceptualization, Methodology and Investigation for Study 1, Writing - Review \& Editing,  Funding acquisition, and Supervision}

\address[4]{These authors contributed equally to this work.}
\cortext[cor1]{Corresponding author}

\begin{abstract}
In this work, we present (i) a novel approach how artificial intelligence can support in the therapy  for better concentration of children with \ac{adhd} through motivational attention training with a virtual robotic agent and (ii) a development process in which different stakeholders are included with their perspectives.
Therefore, we present three participative approaches to include the perspectives of different stakeholders.
Study I: An online survey was conducted with parents in Germany with the aim of ascertaining whether they would use software to promote their children's attention, what influences their attitude towards using it, and 
what requirements it would have to meet. 
About half of the parents would be willing to use software to promote attention. 
Parents are very concerned about the reward and motivational aspect of the software, as well as privacy. 
Study II: To develop the software as close to practice as possible, one of the developers took part in an intensive training for \ac{adhd} (\ac{adhd} camp) with the aim of testing which of the elements are technically feasible. 
Afterward, a first prototype was presented to clinicians to make further adjustments.
Elements that can be technically implemented from traditional behavioral training for \ac{adhd} are mainly direct feedback via a response cost token system as well as verbal feedback. 
Study III: A first feasibility test was conducted with the end users to check if the system works and if children and adolescents can use it. 
Feedback from the end users was collected for further enhancements and adjustments. 
The usability of the software was very good and the children and adolescents found the system quite attractive and motivating. 
Conclusion: Attentional performance software offers multiple opportunities in the treatment of \ac{adhd} if the system is adapted to the needs of the practitioner and end user. 
This development process requires a lot of time and close interdisciplinary collaboration. 
The potential can be seen in relieving parents in homeschooling situations, supporting children with initial attention problems, and prolonging therapeutic treatment effects.  

\end{abstract}


\begin{highlights}
\item novel approach how artificial intelligence can support in the therapy  for better concentration of children with \acs{adhd} through motivational attention training with a virtual robotic agent
\item innovative, interdisciplinary development process in which different stakeholders are included with their perspectives
\end{highlights}


\maketitle

\section{Introduction}
\subsection{Background \ac{adhd}}
\acf{adhd} is one of the most common neuropsychological developmental disorders in childhood worldwide, leading to significant impairments in general functioning. 
The neuropsychological impairments primarily affect executive functions, particularly the ability to inhibit inappropriate responses and state regulation. 
They also have deficits in motivational processes. 
With a global prevalence of approximately 5\%, \ac{adhd} is one of the most common mental disorders in children and adolescents \citep{polanczyk2014adhd}. 
It is often diagnosed before the age of six \citep{schlack2014has}. 
The three main symptoms according to DSM-5 and ICD-10 (International Classification of Diseases (ICD)) are symptoms of inattention, hyperactivity, and/or impulsivity that occur in different situations and have various negative effects, such as on school performance, social functioning, peer relationships \citep{booster2012functional}, and family life \citep{caci2014daily}. Nearly two-thirds of individuals with \ac{adhd} develop a chronic course, meaning that symptoms often persist across the lifespan.

\subsection{(Every day) problems in various areas of life}
The problems associated with \acf{adhd} often become apparent in the early school years. 
There, increasing demands are made on the ability to concentrate and motor rest (e.g., sitting still in class, doing schoolwork at home), which the children can hardly fulfill due to their impairment of psychosocial and cognitive functions \citep{banaschewski2017attention, polanczyk2014adhd}. 
A lack of reward deferral ability is frequently observed, which is the ability to resist the temptation for an immediate reward and wait for a later reward. 
Children with \ac{adhd} often show negative affect ("delay aversion") when experiencing reward delay and try to avoid corresponding situations. 
In this context, attentional and control processes are significantly influenced by the underlying motivation of the child \citep{schmidt2012adhs}.
Furthermore, there are difficulties in action planning (selection of efficient strategies for problem-solving, adaptation, and monitoring of the action). 
Children with \ac{adhd} need more feedback and external support and motivation from the social environment than healthy children in order not to choose dysfunctional solution strategies (e.g., termination, avoidance) in a situation with increased demand for performance \citep{schmidt2012adhs}. 
If tasks are boring or are not supervised, the attention span of children with \ac{adhd} is very limited. 
Adding external stimuli to a potentially boring task can help children with \ac{adhd} to be more motivated to work and to show better performance again \citep{luman2005impact, sader2022rewards}. 
It has already been established that this is not just a matter of pure motivation, for example by giving a simple reward situation. 
In children with \ac{adhd}, so-called response-cost interventions improved accuracy in a math task compared to a simple reward and led to higher motivation \citep{carlson2000effects}. 
In response-cost interventions, children receive direct feedback on demonstrated behavior, for example, by receiving rewards for appropriate behavior and withdrawing them for inappropriate behavior. 
This behavior therapy intervention strategy is used to train cognitive functions, improve inhibition performance, and have a motivational effect \citep{sader2022rewards}.

\subsection{Intensive therapeutic treatment approaches}
Treatment of \ac{adhd} is usually multimodal and integrates behavioral and cognitive treatments, psychosocial interventions (with parents, school), and pharmacological treatment~\cite{somma2019software}.
Therapy for children with \ac{adhd} usually takes quite a long time, which is why researchers have looked at developing a time-efficient, intensive therapy approach: the \ac{adhd} camp \citep{schmidt2012adhs}. 
This is a standardized program that is conducted full-time over a two-week period. 
A fundamental component of the group program is the use of the response cost token system (in terms of contingency management) mentioned earlier for behavior management. 
Children can collect and also lose tokens throughout the day during the different components of the training. 
This occurs as a function of adherence to predefined rules (e.g., focused work, prosocial behavior, inhibition control). 
Of particular note is that the children receive the tokens as promptly as possible as behavioral feedback. Children with \ac{adhd} in particular benefit from such an approach, as it gives them the best opportunity to change their behavior. 
Long-term studies have indicated that children also showed improvement at the functional as well as behavioral level up to a 2 year-period. 
Furthermore, the results of the studies indicate that strategies of instrumental learning through the application of a response cost token system lead to substantial improvements in neuropsychological functioning of children and adolescents with \ac{adhd} \citep{gerber2009adhs, gerber2012impact, kinne2015langzeitwirkungen, sotnikova2012long, toussaint2011wirksamkeit}.

One limitation of the \ac{adhd} camp is that it requires a lot of staff (about 6 to 7 people) to run. 
This is rarely available in both inpatient and outpatient settings, which is why such intensive therapy settings can rarely be offered. 
One question that arises in the context of digitalization and technologization is whether there could be technological solutions to support such promising therapy concepts or also to prolong their therapy effects. 
Various studies on the use of robots have shown that they are certainly capable of providing motivating feedback and supporting people in the implementation of unpleasant tasks \citep{rohm2020persuasive}.

\subsection{State of the art}
There is an increasing market of apps and digital therapeutics for children and young people with \ac{adhd}.
\cite{benyakorn2016current} reviewed current evidence-based technology for \ac{adhd} patients and used the \ac{rdoc} to discuss the potential use of further technology,   which was not explicitly developed for children with \ac{adhd}. 
The six \ac{adhd} \ac{rdoc} constructs are (1) reward-related processing, (2) inhibition, (3) sustain attention, (4) timing, (5) arousal, and (6) emotional lability. 
The authors proposed a theoretical model for implementation and evaluation of technological interventions for children with \ac{adhd} consisting of a recommendation of three components for the development of \ac{adhd} technology: (1) set schedules (2) difficulty matching and (3) immediate feedback.
\cite{benyakorn2016current} analyzed ten existing technologies. 
They argue that although immediate feedback is so important for children with \ac{adhd} “there are no gamified programs specifically designed for \ac{adhd} behavioral modification and no studies have evaluated the effectiveness of gamified programs, but there are many behavioral reinforcement apps that we hypothesize will be effective at increasing motivation in \ac{adhd}”.
Regarding available \ac{adhd} apps, \cite{powell2017attention} stated that research is missing on whether apps are specifically suited for children and young people with \ac{adhd}, and what are the respective key properties for apps to be suitable. 
They interviewed five clinicians and five children and young people with \ac{adhd} to explore their opinions of ten \ac{adhd} apps. They identified five themes, which are important for the children with \ac{adhd} and clinicians who work with them: (1) the accessibility of the technology,  (2) the importance of relating to apps,  (3) addressing \ac{adhd} symptoms and related difficulties,  (4) age appropriateness, and (5) app interaction.
However, they did not include the parents' point of view. 
The requirements that parents place on such an app should already be considered in the development process, as they decide whether their children use it or not.
Meanwhile, research in the domain of gamified technology and serious games for \ac{adhd} treatment has picked up speed. 
Most video games developed so far for \ac{adhd} treatment are computer-based, with a minority of approaches using other platforms such as console or tablet \citep{penuelas2020video}. 
They do not focus specifically on attention but on multiple facets of executive functioning and were only evaluated by affected persons and not by other stakeholders (ibid.). 
Efficacy evaluation is generally lacking high-quality studies \citep{guan2020updates}. 
In 2020, the Food and Drug Administration in the US has approved the first prescription video game treatment for children with \ac{adhd} \citep{canady2020fda}. 
The game called \emph{EndeavorRx} activates the neural system responsible for attention function by presenting personalized sensory stimuli and simultaneous motor challenges. 
Expectation of benefit has been studied in children and parents. 
In a randomized controlled clinical trial treatment effects have been measured using the Test of Variables of Attention (TOVA) and Attention Performance Index (API) which consists of computer-based tasks presenting different visual or auditory stimuli that have to be discriminated by pressing a button, a task similar to the treatment task itself.
\emph{RECOGNeyes} is an interactive eye-tracking game to train the attention control system of \ac{adhd} children, developed by the University of Nottingham \citep{garcia2019novel}. 
Players use their eyes as a game controller and receive continuous, immediate feedback. 
Game tasks increase challenges based on player performance.

In summary, the number of current approaches to \ac{adhd} treatment in children has multiplied in the past few years. 
Most concepts yield the recommendations derived from \ac{rdoc} put forward by Benyakorn and colleagues. 
Novel apps and gamified technology are especially fun and engaging, give immediate feedback and adapt tasks to the current performance of patients.
However, evaluation has been suggested to be of low quality and does not include all relevant stakeholders, in particular patients’ parents are seldom included in the development of technology. 
Very few studies report on the feasibility, user impact or acceptability \citep{valentine2020systematic}.
Additionally, evaluation focuses on similar tasks trained during treatment, and it remains unclear how effects will transfer and treatment will affect real-world situations like doing boring school work.\\
To date, it remains the case that few health care professionals (HCP) are interested in engaging with technology. 
As \citeauthor{valentine2020systematic} aptly summarized, "HCPs' perspectives and the lack of robust evidence base are potentially one of the largest barriers to the wider acceptance broader adoption of technology (p. 13)." 


\section{VACO - Virtual Robotic Agent for Concentration}
Patients with \ac{adhd} have a deficit from delayed reward processing to learn positive or negative consequences (delay aversion due to the deferred reward). 
Hence, they need a timely, immediate reward or negative consequence. In human-human interactions, it could be shown that an intensive, effective behavioral therapy training program for children and adolescents with \ac{adhd} (the \ac{adhd} camp; for a detailed description see \cite{dopfner2004comparative}) has a sustainable effectiveness, particularly on attention regulation, inhibitory control, and executive functions. 
Training is done in a real-world setting with unappealing tasks similar to school work. 
The main feature of the training is an intensive \ac{rct} system~\citep{gerber2012impact}. 
The study by \cite{gerber2009adhs} was able to show that the positive long-term effects of the \ac{adhd} camp could be predominantly attributed to the effect of \ac{rct}.
In the \ac{vaco} project, we investigate how this training can be supported by the virtual agent Flobi. 
The main research questions of this project are (i) Which parts of the training can be implemented with the agent? (ii) Can additional training with the agent at home increase the positive effects of the \ac{adhd} camp?

Children with \ac{adhd} need immediate rewards to shape their behavior. 
Flobi should take up exactly this aspect and reward the children directly for desirable, attentive behavior through a response cost token system. 
In previous research, Flobi already incorporated the attention state of its interlocutor and reacted with non-intrusive intervention strategies to inattentive interaction partners \citep{richter2021attention}. 
The fact that the child does not experience a delay of gratification may lengthen his or her attention span and may reduce unwanted interfering hyperactive impulsive behavior. 
However, while computer games often follow a fixed reward pattern which might become boring, the virtual assistant Flobi can interact with the children. 
This innovative aspect is an essential difference to previous computer-based approaches. 
It is expected that the child after the training will show more self-control.

\subsection{Participatory development process}
Whether innovation within academia finds its way into practice and is later accepted and used depends on many factors. 
\cite{klemme2021multi} developed an academic knowledge transfer strategy for healthcare technology. 
They pointed out that professional exchange between practice partners from healthcare and academic partners is mandatory for generating new research hypotheses. 
This equal participation of the practice partners during the development process allows an agile process, early evaluation of the approach, and human-centered design.
Our development process of the system therefore involves several stakeholders:
The \textbf{clinician team} shows practical needs in which technical support is desired during the \ac{rct} intervention, whereas the \textbf{development team} takes care of the technical implementation and can identify which parts of the \ac{rct} system are technically feasible. 
The system itself is developed in a co-design process. 
Further stakeholders are integrated into the process at an early stage.
\textbf{Parents} are the decision-makers whether such a system is used at all. 
It is therefore important to involve them early in the development process. Therefore, we performed an online-survey with 517 parents of \textbf{children}, addressing the question if they would use a software for attentional training.
In addition, their intention formation for using such a software and their expectations regarding it are addressed.
Besides the clinicians and the parents, the children are the end-user of the software and are one of the first test subjects for evaluating the usability of the first prototype.

In the following, we explore the perspectives of the three stakeholders in more detail.

\section{Parent's perspective: online survey regarding expectations}
Parents are one of the decision-makers whether such a system is used at all. It is important to involve them early in the development process. We therefore performed an online survey to answer the following research questions:
\begin{enumerate}
\item[] \textbf{RQ Usage:} Would parents (especially in the stressful situation of homeschooling) use a software that could recognize the child's attention and intervene in a motivating manner?
\item[] \textbf{RQ Influences:} What influences the formation of a parent's intention?
\item[] \textbf{RQ Expectations:} What would they expect from this software, what should be considered during development?
\end{enumerate}

\subsection{Method}\label{sec:study_1_method}
The survey was carried out from November 2020 to mid-February 2021.
During this time, families were heavily burdened by the Corona pandemic, as society was in lockdown and parents had to homeschool their children independently. 
This often created a double burden of work (sometimes in the home office) and teaching. 
In many families, the level of stress increased and conflicts arose more frequently, especially between parents and their children
Since this is a particularly unusual time, we also addressed the parents' experience of stress to determine the influence on the willingness to use the software.
Parents were asked about their subjective experience of stress, their children's need for support, and their interest in support software for concentration promotion. Parents of children in the first to 13th grade in Germany asked about the special situation of homeschooling in the corona situation. 
The following survey instruments are used:\\
\textbf{Demographics:} General demographics are assessed, including age, gender, and type of school of the children (8 items).\\
\textbf{Media usage behavior:}  (4 items).\\
\textbf{Parental stress:} To access the parental stress, the parental stress questionnaire by \cite{domsch2010esf} is adapted to the current corona pandemic situation (12 items).\\
\textbf{Children's need for support:} Parents estimate the frequency with which their children need support (3 items) and the time they need help (3 items).\\
\textbf{Interest in motivational software:} After a brief introduction of the learning software “Flobi” to support their children's motivation, parents were asked if they would use such an app (1 item). 
To get a more differentiated picture of their intention formation, they are additionally asked about the attitude (4 items), subjective norm (3 items), perceived behavioral control (4 items), behavioral intention (2 items), and their past behavior regarding learning software (1 item). 
These items were constructed following the \ac{tpb} \citep{ajzen1991theory} that claims to show which components influence the formation of a person's intention \citep{ajzen2015theory}.\\
\textbf{Expectations from this software:} In a free form text, parents had the opportunity to express themselves what they would expect from this software (see \autoref{app:parent_survey}).
\\
For the analysis, three experts create cluster categories from the free form texts. Two additional persons annotate each answer to one or more of the categories. In case of disagreement, the experts discuss it and decide together.

\subsection{Results}
\subsubsection{Participants}
In total, 517 parents of children in the first to 13th grade in Germany take part in the study. 
Most of the children go to high school (61\%), while a third went to primary school (32\%). 
In their self-assessment, around 89\% of parents rated their own media skills as high to very high.
Detailed results of the parents' subjective experience of stress and the children's need for support can be found in \cite{petras2021familien}. 
It has been shown, that the stress experience in homeschooling in the corona pandemic is high, especially for parents of younger children. 
An important factor is the frequency with which parents have to keep motivating their children to start or continue with schoolwork. 
A large proportion of parents (68\% of parents with children in grades 1-4) stated that it was stressful to have to motivate the child to do their schoolwork several times.

\subsubsection{Usage}
The first research question addresses whether parents would (especially in the stressful situation of homeschooling) use a software that could recognize the child's attention and intervene in a motivating manner.
13 participants did not provide any information on whether they would use such supporting software. 
Of the remaining 504 participants, 165 participants (32.7\%) stated that they wanted to use the software, 159 (31.5\%) don't want it, and 180 parents (35.7\%) are unsure.

\begin{figure}
    \centering
    \includegraphics[width=0.9\linewidth]{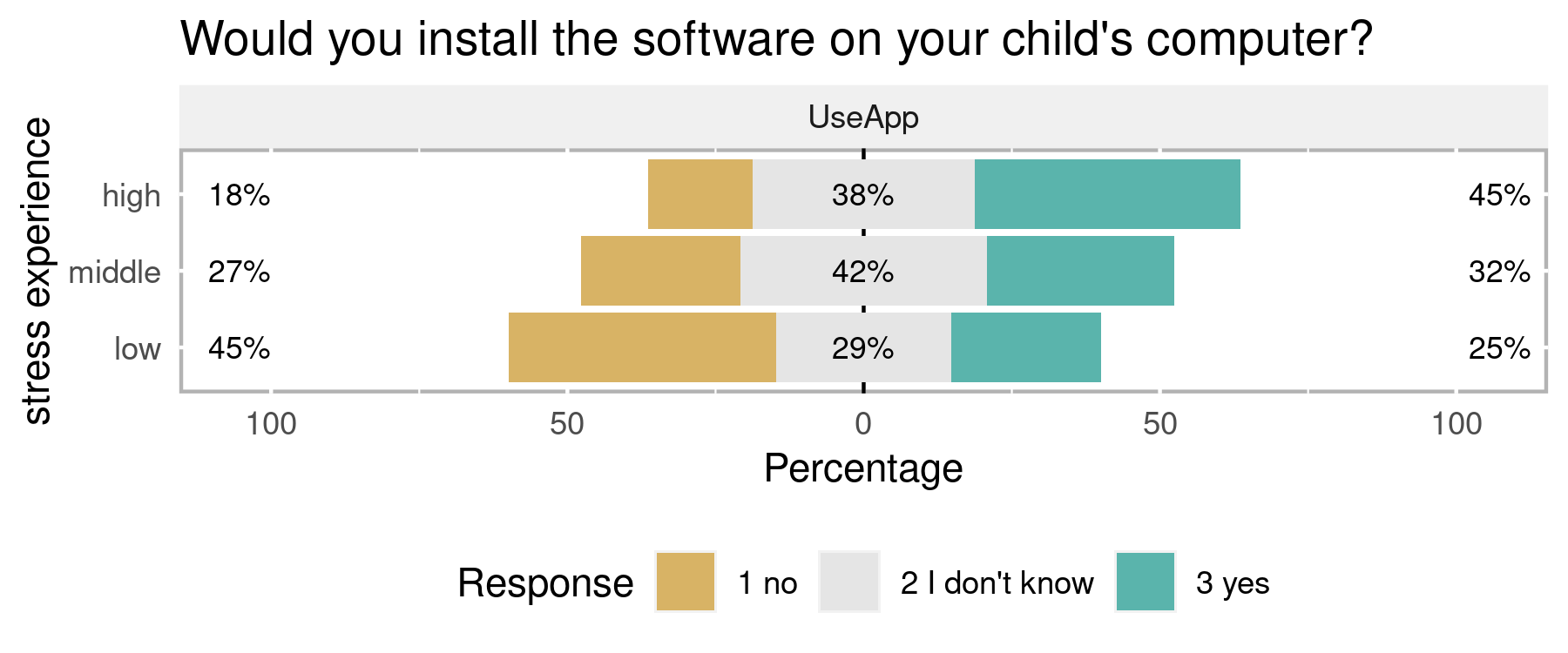}
    \caption{Parents' interest in the motivational software, depending on their stress experience.}
    \label{fig:parents_usage}
\end{figure}

\autoref{fig:parents_usage} depicts the results of this question depending on the parent's stress experience (low, middle, high). 
A chi-square test was conducted between the stress experience and whether the software would be used. 
No expected cell counts were less than 5. 
There is a statistically significant low association between the stress experience and the willingness to use a motivational software, X²(4) =34.43, p < .001, V = 0.15. 
Parents with a high level of stress would be more likely to install the software than parents with a lower stress experience.

\subsubsection{Influences}
The second research question addresses what influences the formation of a parent's intention.
\begin{figure}
    \centering
    \includegraphics[width=\linewidth]{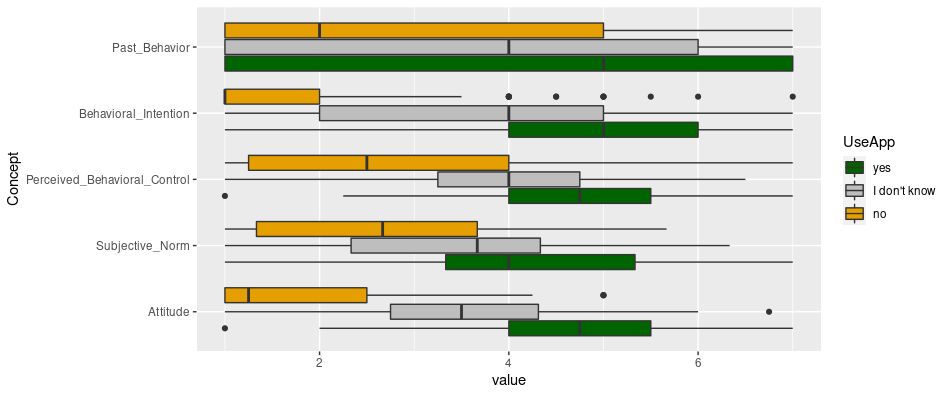}
    \caption{Influence of the\ac{tpb} concepts on the parent's intention to use the software.}
    \label{fig:parents_TBP}
\end{figure}
To get a more differentiated picture of their intention formation, the results of each component of the \ac{tpb} is presented shortly (see \autoref{fig:parents_TBP}).
All TPB-items were assessed using 7-point Likert-scales, with item responses ranging
from 1 (strongly disagree) to 7 (strongly agree).
\\                   
\textbf{Behavioral intention} (2 items, Cronbach's $\alpha = .96$):\\
There is a statistically significant moderate association between the behavioral intention to use the software and the willingness to install it on the children's computer, X²(24) =278, p < .001, V = 42. The average behavioral intention to use the software is 3.40 (SD=2.02).
As expected, parents who would use the software have a higher willingness to install it.\\
\textbf{Attitude towards the software} (4 items, Cronbach's $\alpha = .93$):\\
The average attitude towards the software is also very divided (M=3.38; SD=1.66) and it exists a moderate statistically significant association between the intention to use the software and the attitude towards it, X²(48) = 304.84, p < .001, V = .42. As expected, parents who would use the software have a higher positive attitude towards it.\\
\textbf{Perceived behavioral control} (4 items, Cronbach’s $\alpha = .73$):\\
The average perceived behavioral control was rated quite high (M=3.89; SD=1.51) and also a moderate statistically significant association between the intention to use the software and the perceived behavioral control exists, X²(48) = 221.33, p < .001, V=.38. Parents who would use the software are more likely to believe that their children also want to use it and can handle it.\\
\textbf{Subjective norm} (3 items, Cronbach's $\alpha = .76$):\\
Furthermore, a moderate statistically significant association between the subjective norm and the intention to use the software exists, X²(38) = 153.16, p < .001, V=.31. On average, parents rated the subjective norm with 3.58 (SD=1.50).
Parents who would use the software are more likely to believe that their acquaintances would like it if their child could use such an educational software.\\
\textbf{Past behavior regarding learning software} (1 item):\\
Interestingly, there is only a statistically significant low association between the past behavior regarding learning software and the intention to use the new software, X²(12I)= 50.558, p < .001, V=.18. The past behavior received on average 3.74 (SD=2.49). Parents who would use the software were more likely to use educational software in the past

\subsubsection{Expectations}
The last question addresses what parents expect from the software and what should be considered during development.
The parents had additionally the opportunity to express their expectations regarding such software and were clustered by three experts (as explained in section \ref{sec:study_1_method}). 
The following figure visualizes the classifications of their expectations and how often each of these categories occurred, depending on whether the software would be used or not.

\begin{figure}
    \centering
    \includegraphics[width=\linewidth]{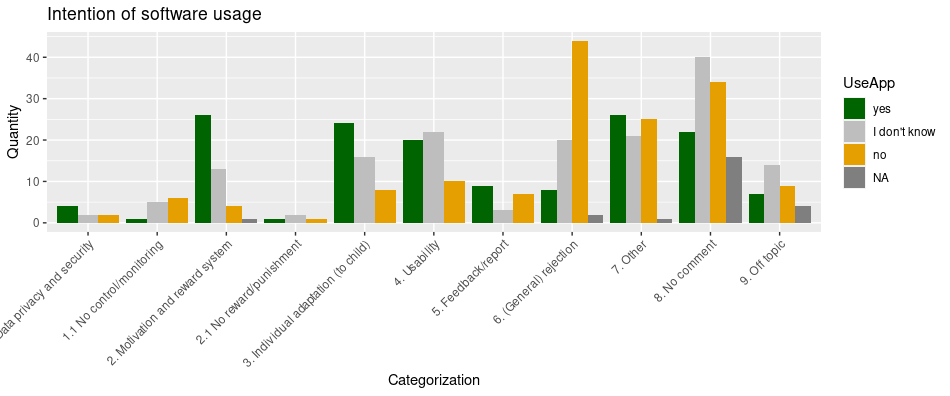}
    \caption{Classifications of parant's  expectations regarding the software.}
    \label{fig:parents_Expectations}
\end{figure}

\textbf{Data privacy and security} was mentioned 14 times. Parents are primarily concerned with general data security for the child, e.g., “100\% safety, protection of the child from external influence […]” [VP143]. 
19 parents noted that they don't want any \textbf{monitoring}. One parent indicated that they want "motivation, not control" [VP381].
The \textbf{motivation and reward system} was addressed 82 times, especially by parents who want to use the software. 
They expect, for example, ”exactly what was described above: to motivate, but also to show the limits of resilience” [VP406], or that the software “promote concentration” [VP339]. 
However, seven parents stated that they wish \textbf{no reward or punishment}. 
They generally see a point system as critical, e.g. “children should not be trained to learn like in a game, to collect points and rewards.” [VP508].

An important expectation on the software is the possibility of \textbf{adaptation and individualization} to the child, which was addressed 77 times. 
The software should present “exercises that adapt to the level of the child” [VP386] and should be “age-appropriate” [243]. 
The adaption of the exercises to the individual level of the child should be possible either automatically by the software or externally (e.g. by the teacher).
Furthermore, the main expectation was aimed at the \textbf{usability of the system} and was addressed by 86 parents. 
It is important to the parents that the software is “intuitive to use [...]” [VP88] and the “[...] child should be able to work with it independently and success-oriented [...]” [VP98].

In addition, 23 parents suggested some kind of \textbf{feedback and report} for the children, but additionally to discuss it with other (e.g. the teacher). 
One parent stated, “There should be direct feedback to the students (audio and visual) and a report made over a few days that could be used to discuss with older students how much time they have been concentrating and how much time was 'break'.” [VP461].
85 parents have not formulated any expectations because they generally reject the system. 
Some parents believe that the system will not work (e.g., “None at all, because it won't work” [VP451]) and several other parents mentioned that their child does not need this software (e.g. “None - I am lucky that my child does everything independently and conscientiously without motivation![...]” [VP469]).
84 additional comments could not be assigned to one of the categories and marked as \textbf{other}. 
This includes suggestions for additional features, e.g. enabling video conferencing with teachers or suggesting sports exercises. 
Furthermore, in includes general remarks regarding the age of the target group.
112 parents left \textbf{no comments} at all, and 53 parents had additional comments on the current homeschooling situation \textbf{(off-topic)}.

\subsection{Discussion}

The main expectation was aimed at the \textbf{usability of the system} and should therefore be one of the priorities during implementation. 
Furthermore, \textbf{motivational aspects} were discussed frequently. 
Several parents want a motivating aspect, but without punishment. 
This does not consider the fact that rapid feedback on  inappropriate behavior  is particularly important for children with \ac{adhd}. 
However, the \ac{rct} system provides a good possibility to integrate these motivational aspects.
In addition, several studies have demonstrated the effectiveness of \ac{rct}s in the treatment of \ac{adhd}.

The \textbf{individual adaptation} was another important point for the parents and is something that current systems rarely implement. 
In the current system, this should be implemented in two ways. 
On the one hand, the task itself should adapt to the age. 
The vision is that the system will recognize a user's strengths and weaknesses and adjust the task difficulty accordingly. 
This would be especially helpful in supporting school assignments, as it is not always possible for teachers in regular classes to address the individual needs of students.
On the other hand, the system’s motivation should be based on the attention of the user. 
Few parents wish some \textbf{report} for the parents/teacher about the individual progress of the children. 
This conflicts somewhat with the desired data security and the \textbf{undesirable monitoring} of the children.
However, the \textbf{feedback} for the children itself is the most important part of the \ac{rct} system. 
An additional report for other people (e.g., parents, teacher, or therapist) is conceivable and could be created with the consent of the child.

Some parents expressed a \textbf{general rejection} of further use of software (less screen time), especially during homeschooling/corona. 
This finding fits our assumption that not all children benefit from the software, but that indicated use would be best. 
Especially technology-savvy children who already developed an initial dislike of homework or school could be supported and motivated by the software. 
This in turn could have a positive effect on how much they enjoy going to school.


\section{Clinician's and developer’s perspective: the first prototype of the system}
To decide which parts of the training can be implemented with the agent, the developer visits a \ac{adhd} camp to get a better understanding of the therapy training program for children. 
Afterward, a first prototype was developed in close cooperation with the clinicians.
However, due to the Corona situation, discussions about the systems were mostly conducted online. 
Before being tested with end users, the system was presented at KJP's premises and tested by clinicians. 
This should identify major issues before testing them with end users.

\subsection{Material}

\begin{figure}[ht]
    \centering
    \subfigure[Setup]{\includegraphics[width=0.35\textwidth]{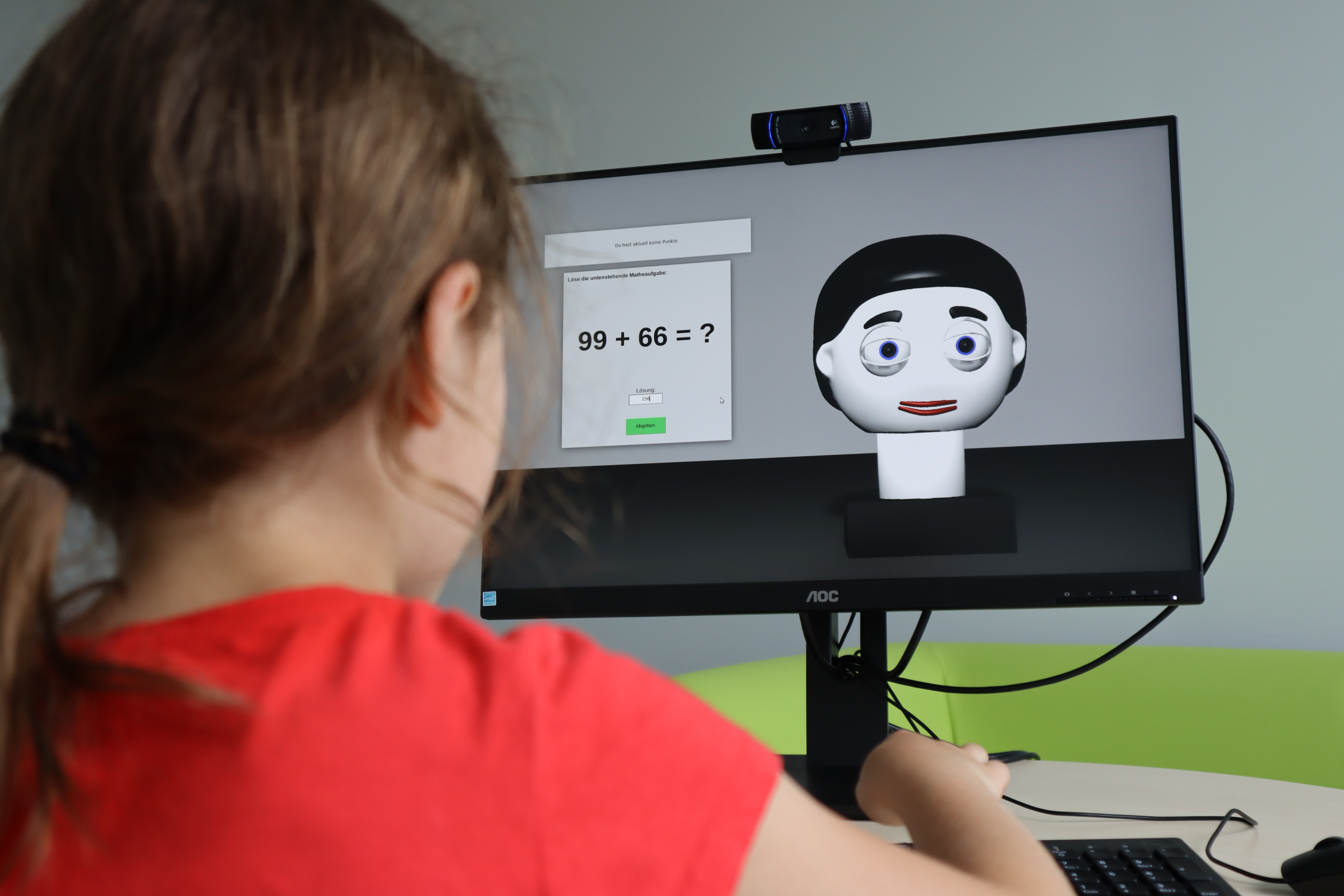}}    
    \subfigure[Gaze tracker]{\includegraphics[width=0.5\textwidth]{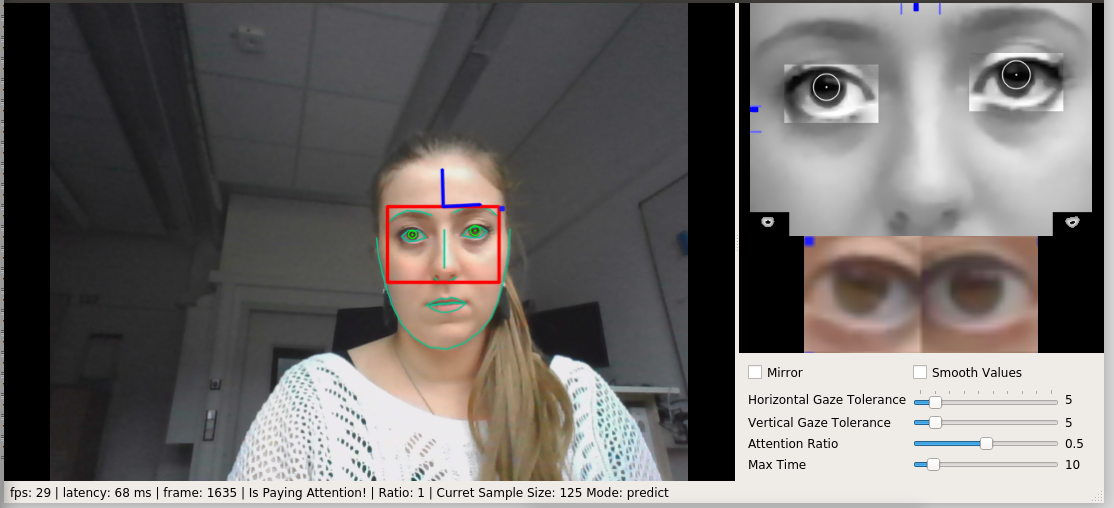}}

\caption{(a) The setup: children interact with the virtual agent Floka. (b) Visualization of the gaze tracker.}
\label{fig:vaco}
\end{figure}
The clinician-developer team decided to start with the school situation of the \ac{adhd} camp. 
In this situation, the children perform some math tasks, which should be monotonous as possible, to specifically address concentration performance. 
The children's task is to look at their worksheet and work with concentration. 
In doing so, they should not be distracted by distracting stimuli (e.g., conversations of others, drinking bottles falling over). 
These distracting stimuli are actively presented by the trainers and intensified over the course of the training sessions. 
If a child is distracted, a point is deducted and the inappropriate behavior is specifically named. 
If the child then returns to the task and continues to work in a concentrated manner, he or she again receives direct verbal behavioral feedback and a point.
\autoref{fig:vaco} shows the implementation of the school situation with the virtual agent.
\\
\textbf{Task:} On the left side, the simple math tasks are presented. 
These are adaptable to the age of the children. 
Furthermore, it gives intermediate feedback for the task. 
A new task will only appear once it has been solved correctly (which, however, is not decisive for whether the child receives a point or not). 
Additionally, if the result submitted by the child is wrong, a short feedback sound is provided.
\\
\textbf{\ac{rct} system:} On the right side of the screen, the agent is visualized. 
It provides positive and negative feedback on the attention state of the child according to the principles of the \ac{rct} system.
Positive feedback will be provided when the child starts working quickly (\textbf{praise\_immediate\_start}). 
This is measured through fast key input after the task explanation by the agent.
Furthermore, it provides positive feedback when the visual attention of the child is on the screen, e.g., “You're doing great. Another point for you.” \textbf{(praise)}.
Flobi's AI decides based on an SVM whether the children are attentive or not. This is measured via the webcam on top of the screen.
A further development of a gaze detector~\cite{Schillingmann2015}, based on the  dlib's face alignment model~\cite{king2009dlib}, is used to decide whether the child is looking attentively at the screen or "looking around". 
Accordingly, negative feedback is provided whenever the child’s visual attention moves away from the screen or the keyboard \textbf{(criticize)}, e.g., “You are inattentive, try to concentrate on the tasks again!”. 
The feedback itself is multimodal: verbal and facial expressions by the agent, and a visualization of the points presented over the task. 
In addition, the agent can try to distract the child by making faces or verbalizing short phrases like “oh, look behind you”.
\\
\textbf{Individualization possibility by clinicians:} The system provides a small visualization for the clinicians. 
This allows to load a specific configuration for a child, including the following setting options:
\begin{itemize}
\item name of the child
\item age of the child: for the adaption of the math tasks
\item child ID: for the generation of the report
\item session ID: for an adaption of the feedback quantity over time
\item degree of distraction: adaption of the quantity of distraction depending on the child's ability to concentrate.
\end{itemize}
Based on this setting, the \ac{rct} system will adapt its behavior. Furthermore, the interaction can be started through the visualization.

\subsection{Method}
The system was presented in January 2022 at the KJP by one of the developers. Afterward, the clinicians can take on the role of the children and test and try out the system themselves. 
The clinicians consisted of the head of the university hospital, who is a specialist in child and adolescent psychiatry as well as a professor of child and adolescent psychiatry, and a psychologist who is involved in both inpatient treatment and research.

During the exploration by clinicians, several implementation decisions regarding the design of the software and the resulting behavior are discussed. In the following, the major results affecting the first usability test with the end users are presented shortly.

\subsection{Results and Discussion}
\begin{figure}
    \centering
    \subfigure[Floka simulation]{\includegraphics[width=0.3\textwidth]{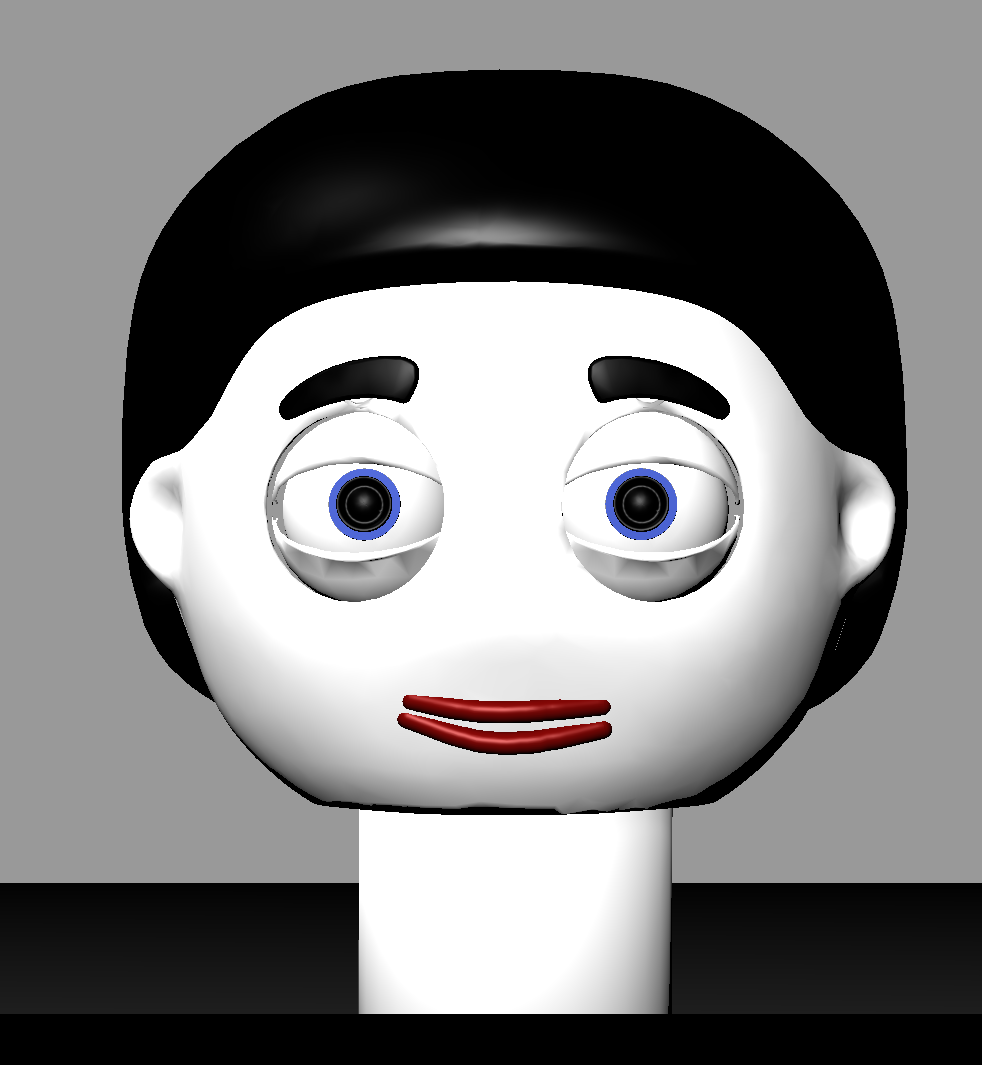}}
    \subfigure[Flobi simulation]{\includegraphics[width=0.3\textwidth]{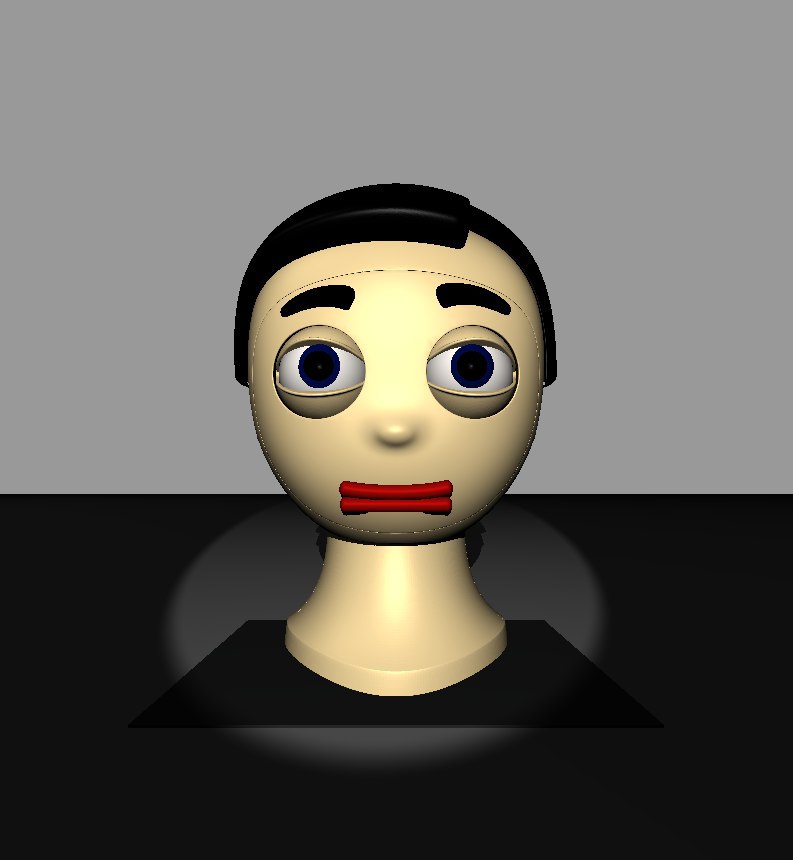}}
\caption{The two virtual simulations of the robot head. On the left side is Floka, on the right is the agent Flobi.}
\label{fig:flokaFlobi}
\end{figure}
Several design decisions are discussed and adapted. The main points are presented in the following.\\
\textbf{Resulting design changes:}  The system can be represented by two different versions of the agent Flobi, depicted in the figure \ref{fig:flokaFlobi}. 
On the right, Flobi is presented and on the left the subsequent version called Floka. 
The system was designed with the Floka version. 
However, the clinicians preferred the Flobi visualization because it was perceived as more child-friendly. 
Furthermore, the agent’s position was discussed. 
It should be slightly further to the right and smaller. 
In relation to the task, it should not be overwhelming. 
In addition, the default facial emotional expression of the agent should be changed from neutral to slightly happy.
\\
\textbf{Resulting task changes:} The system provides intermediate feedback for the results of the tasks. Similar to the short sound feedback by submitting a wrong answer, a short positive sound feedback for answering the question correctly should be added.
Furthermore, since math tasks often cause evaluation anxiety, other task types were discussed, e.g., performing more \ac{adhd}-specific attention tasks.
\\
\textbf{Resulting \ac{rct} sytem changes:} During the discussion, the need for additional feedback classes arose.  Shorter reward phrases with a direct salutation for attention should be integrated, e.g., “Keep it up, <NAME>!” (\textbf{short\_praise}).
Additional to the normal positive feedback, a special praise after the child is attentive after an attention lost would be useful (\textbf{praise\_after\_reattention}), e.g., “It's great that you're concentrating again!”. 
This is a positive confirmation for the behavior change after a criticism. 
In contrast, after 30 additional seconds of inattention, points are deducted again, verbalized, e.g., “Unfortunately, you're still distracted.” (\textbf{criticize\_again}).

\section{Children's perspective: first usability test with end-user}
For the first usability test with end-user of the prototype, the resulting behavior and design changes have been implemented. The  pilot test itself was carried out again in the KJP. The goal was to test the general usability of the system, to get a first feedback from end users, and by this  to identify the potential for improvements. 

\subsection{Material and Method}
The usability test was performed in March 2022. 
The system was set up in one of the rooms in the KJP. The parents of the children were informed about the study in advance and gave their consent for their children to participate. 
Additionally, the children were informed by a clinician and also gave their consent for the study.
Each child interacted in two trials each with the system. 
The first trial started with a brief introduction about the system and the task itself by Flobi. 
Afterward, participants had to solve the math task for five minutes. 
During the task processing, Flobi gave multi-modal feedback on the child's attention state. 
After a short break, the children started a second trial. 
During this, Flobi additionally tried to distract the children two times with short phrases.
At the end, Flobi said goodbye to the children.

After the interaction, we conducted a semi-structured interview (see \autoref{app:children_survey}), addressing the usability of the system and finding potential for improvement. 
Therefore, the \ac{sus} \citep{brooke1996sus} was converted into simple language with a clinical expert. 
In addition, four questions about the points system and Flobi were asked.
The participants were able to answer using a 5-point likert scale, which was presented by a visual analog scale for practical work with children and adolescents \citep{grasser2017rating}.
The interview ended with open-ended questions about opportunities for improvement.

During the interaction, the videos of webcam (Flobi’s view), the screens and several system events were recorded. 
Furthermore, the audio of the interviews were recorded and transcribed for later analysis.  
The study was approved by the ethics committee of Bielefeld University.

\subsection{Results}
\subsubsection{Participants}

In total, ten children (8 female, 2 male) interacted with the system. 
They were recruited from the adolescent ward (n=4), pediatric ward (n=3), and day hospital (n=3) of the child and adolescent psychiatry. 
No diagnosis of the children were obtained, since these are irrelevant for the question of our feasibility study.
The age range from 8 to 17 with an average age of 12 (SD=4.99). Participant VP07 is excluded from the later analysis. 
She dropped out because she was overwhelmed with the situation due to her own stress and felt uncomfortable removing her FFP-2 mask (Corona protective measure) in the presence of strangers.

\subsubsection{System Usability Scale}  
\begin{wrapfigure}{R}{0.7\linewidth}    
\centering
    \includegraphics[width=\linewidth]{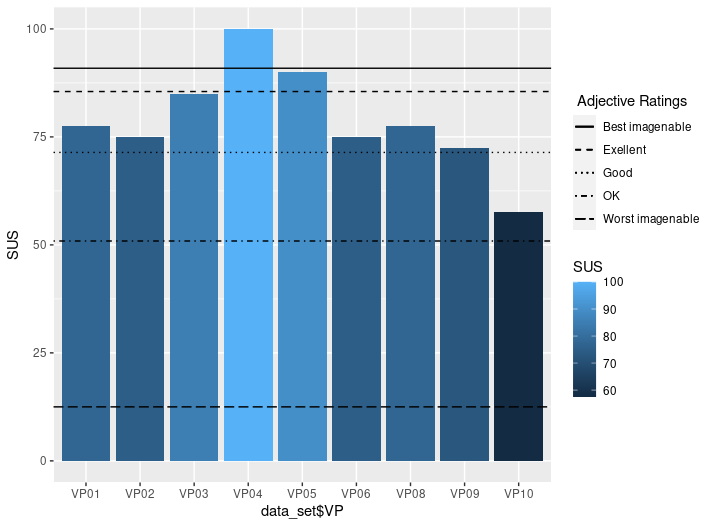}
    \caption{The system usability scale for each participant.}
    \label{fig:children_SUS}
\end{wrapfigure}
The following figure visualizes the results of the System of rules \ac{sus} for each participant (except for VP7, due to interaction abort). 
On average, the system received a \ac{sus} value of 78.89 (SD=11.93), which can be interpreted as “good” \citep{bangor2009determining}. 
One child rated the usability as “excellent” and another one even as “best imaginable”. 
Participant VP10 rated the usability only as “OK”.

\subsubsection{Questions regarding Flobi and the scoring system}
The following figure depicts the results of the four questions regarding Flobi itself and the scoring system. 
Most children (n=8) \textbf{understood the praising or advice} regarding their attention. 
Only participant VP10 wasn't sure about this. 
Furthermore, most children found \textbf{Flobi to be fair} (n=5), three children were unsure about that and participant VP10 found Flobi not fair. 
Five children \textbf{liked the scoring system}, two children were unsure about it, and two children rather disliked the scoring system. 
Regarding the \textbf{motivational} aspect, again, five children found Flobi motivational. 
One participant was unsure, and three children found Flobi (rather) not motivational.

\subsubsection{Open questions regarding the motivational program and Flobi itself}
Two children indicated that they would not change anything in the program. The other seven participants suggested the following points for improvement.
Four children suggested improvements regarding the \textbf{fairness} of Flobi, e.g., VP09 stated “[...] Thinking or looking at the keyboard was considered a distraction”.
Indeed, the eye-tracker detector had some difficulties detecting the face when the children looked down at the keyboard, especially with some hairstyles where the hair covers the eyes while looking down. 
However, looking around–also for thinking about the math tasks–is counted as “undesirable behavior” during the \ac{adhd} camp and should be avoided.

Three children commented negatively on the \textbf{feedback} itself regarding their attention. 
One child found the personal attribution with their names as “rather annoying”, and the other two found the feedback too frequent, although praise was well received.
Regarding the \textbf{feedback of the task itself}, two children commented that they found the feedback sound annoying.

Further potential for improvement offers \textbf{Flobi's appearance}. 
Three children would improve Flobi’s lips, e.g., VP05 stated, “definitely the lips, so they look a bit more human and not just two bars.[...]”. 
Flobi’s eyes were addressed three times, e.g., “Being able to see through the eyes was a little weird.[VP09]”. 
This is because only the outer shell of the robot is modeled in the simulation. 
Since Flobi itself has a modular structure, the individual parts in the simulation can be seen. 
Two children perceived this as “cracks in the skin”. 
Furthermore, two children want a more feminine Flobi. 
One child found the pitch of Flobi's voice tiring.

\subsection{Discussion}
The aim of this first pilot test was to find out 1) whether the system works in use with children and adolescents, 2) whether the children accept the system and Flobi, and 3) whether the feedback from Flobi as well as the integrated \ac{rct} system has a motivating effect.

The results indicate that all participants, regardless of age, were able to use the system well (even the 8-year-old), which is reflected in the high usability values. Furthermore, younger participants found Flobi similarly appealing as the older participants. 
This contrasts with the parental assessment that only younger children might find the system appealing (see above parent's perspective). 
It is important to note, of course, that this was a small sample and artificial laboratory conditions, and actual, longer-term application in everyday life is necessary to verify target group suitability.

In terms of applicability, it became clear that difficulties can arise when children do not adopt the "correct" sitting position or have hairstyles that obscure the face. 
One possible solution to this problem is to use newer software that is less prone to failure in this regard. 
In general, the system and Flobi were well received by most participants. Only the external appearance of Flobi was criticized more often.
Whether Flobi has a motivating effect on the children can be confirmed in half of the participants. 
This is a very positive finding for further research, which in the next steps can look at which patients Flobi has the most motivating effect on. 
Here, for example, it is important to look at different psychiatric disorders and also to focus on its use with children and adolescents with \ac{adhd}. 
The same applies to the \ac{rct} system, which was also perceived positively by half of the children. 
The results further illustrate how sensitive the children are to a feedback system and how important a properly functioning software is, as the children's focus is very much on the equity aspect of the feedback. 
This aspect is likely to be particularly important in future research with children with \ac{adhd}, as these children often feel that they are being treated unfairly and have a heightened sense of justice. 
At the same time, this also illustrates once again the potential in terms of the effectiveness of the system and of Flobi, as these children may feel particularly addressed and motivated by an appropriately well-functioning, fair software.  


\section{Conclusions and Future Work}
The overall goal of the research group is to develop a digital attention training with the virtual agent Flobi, which should primarily support the treatment of children and adolescents with \ac{adhd}. 
To this end, an interdisciplinary team has joined forces, combining the neighboring disciplines of psychology, psychiatry, medicine, engineering and computer science. 
After creating the necessary conducive conditions for a successful collaboration in a first work phase (understanding of the respective other discipline, research models, common goal definition, etc. cf. \citep{brandstadter1medizintechnik}), the corresponding hardware and software were to be developed in a participatory process.
This article presents the joint approach and aims to answer three major questions. 

First, a general online survey indicated that parents feel overwhelmed and stressed in a homeschooling situation and would be quite willing to use software to help their children focus on homework. 
The need for a corresponding development was therefore given. 
To adapt the corresponding system as well as possible to the requirements of the children and adolescents, the relevant experts were closely involved in the development process (sub-study II). 
This participatory approach proved to be very beneficial, as the subsequent practical test with the children and adolescents was very positive. 
They were able to get involved with the system, were motivated by Flobi and the \ac{rct} system, and rated the system as thoroughly positive overall. 
This is a strength of our approach, as we chose a multi-stakeholder development process and evaluation. 
We collected the parents' perspective and included it in the development. 
For example, usability was very important to parents, which we successfully implemented. 
Equally important was the possibility of individual adaptation, which is also given through adaptable task difficulty. 
Furthermore, the suggestions and criticisms of the clinicians were taken up, whereby the implementation principles of the \ac{rct} are addressed. 
And finally, the implementation was reflected with the children and their feedback was included in the further revision of the system. 
This kind of development of a hardware and software adapted to the needs of the end users is a unique selling point in the child and adolescent psychiatric field, and absolutely necessary to ensure engagement with the technology \citep{valentine2020systematic}.
Frequently, ready-made products are sold that are of little use in everyday clinical practice. 
We want to avoid this, and for this, we accept a longer-term development process.

We have presented a novel approach to how artificial intelligence can assist in the therapy of children with \ac{adhd} through attention training with a virtual robot agent. 
The development of the software builds on an already established and evaluated training (\ac{adhd} camp) and integrates these principles into a digital solution approach. 
Unlike other digital \ac{adhd} trainings, Flobi gives intermediate positive and negative feedback that implements some kind of interaction with the user. Furthermore, the system is adaptable to children in terms of task difficulty and attention feedback.

\subsection{Limitations}
The reported results and conclusions must be viewed critically. 
Firstly, not only parents whose children were diagnosed with \ac{adhd} were interviewed. 
It could be that this specific target group has different needs and different expectations from software than in the general population. 
At the same time, the development process is still at a point where the final direction is still to be determined.
The program may be better suitable in the context of indication-based prevention than for general intervention in \ac{adhd}. 
Furthermore, the sample of both clinicians and end users was small, so its appropriateness and acceptability will only become apparent in future studies with larger samples. 
This includes the fact that end users who explicitly had \ac{adhd} been not selected in this initial phase. 
In addition, only interested children and young people from the clinic took part in the study.
It is unclear how less technically interested end users would rate the usability of the system.
Future studies will also need to work with different control group designs here to see if the software is similarly effective across all target groups or if, for example, children with \ac{adhd} need an adapted version. 
The first results indicate that the software is accepted and can also be handled. Whether the motivating aspect of Flobi is maintained over a longer period of time or whether the software would be used independently cannot be said. 
Again, studies are needed to answer these questions. 
Additionally, the items for the various scales were designed by the authors themselves, which is why the reliability and validity of the scales must be viewed critically.



\subsection{Future work}
Our future research will focus on further optimizing the system by integrating the end users' suggestions for improvement. Furthermore, there will be a software update so that we can hopefully expect a more accurate functioning of the eye tracker in a next test run. Additionally, we would like to extend the task format and integrate, for example, the QB test, which has already been successfully used by other researchers in the digital treatment of \ac{adhd} \citep{hall2016clinical,hollis2017annual}. 
In a next step, the Flobi system will function as an integrative part of the \ac{adhd} camp. 
Here, a first effectiveness study with control group design is targeted. 
One group will participate in the \ac{adhd} camp (treatment as usual), while the second group will receive additional attention training through Flobi. 
If positive results are shown for the use of the system, we want to introduce Flobi in a subsequent study at the  \ac{adhd} camp and then give the program to the families to take home.
An intended question here is whether the integration of the Flobi system into an already long-term effective intensive therapy program \citep{gerber2012impact} with subsequent home-treatment for a certain period of time, can prolong the positive effects of the \ac{adhd} camp. 
Again, a randomized controlled trial with a control groups is sought to ensure high quality of evaluation \citep{somma2019software,valentine2020systematic}. 
Overall, the system is to be permanently further developed in a participatory process. 
It is also conceivable to examine whether the system can also be used as a supporting element in the diagnosis of \ac{adhd} (e.g., using the QB test). 
The visionary end product could be that the Flobi system can be used for the treatment of \ac{adhd} as well as for the general promotion of attention performance in children with first signs of an attention deficit. 
It is conceivable that the system could also be used in a school context. 
For example, it would be a breakthrough if school assignments could be uploaded to the system and the child could then work on them at home.
In doing so, Flobi would provide the child with motivational and attention-grabbing support and, optimally, also recognize when a child is tired, for example, to then encourage a break. 
At the same time, the system could record the child's performance level and individually adjust the difficulty of the tasks.


\newpage
\appendix
\section{Online Survey Regarding Expectations}\label{app:parent_survey}
\textbf{Usage:}\\~
\textit{Would you install software or an app on your child's computer that could recognize your child's attention and intervene appropriately in a motivating manner?}\\
Würden Sie Ihrem Kind auf dem Computer eine Software oder App installieren, die die Aufmerksamkeit Ihres Kindes erkennen könnte und entsprechend motivierend eingreifen würde?\\

\textbf{PE: \textit{Performance Expectancy}}

\begin{enumerate}
    \item \textit{I think my child would benefit from using the educational software ("Flobi").}\\
        Ich denke, mein Kind würde von der Nutzung der Lern-Software („Flobi“) profitieren.
    \item \textit{I think the use of the learning software ("Flobi") is…}\\
        Das Nutzen der Lern-Software („Flobi“) finde ich…
    \item \textit{I think the learning software ("Flobi") would relax the homework situation at home.}\\
        Ich denke, dass durch die Lern-Software („Flobi“) die Hausaufgabensituation zu Hause entspannter werden würde.
    \item \textit{I think that I would argue less with my child about the homework or would not have to discuss so much with the learning software ("Flobi").}\\
        Ich denke, dass ich mich durch die Lern-Software („Flobi“) weniger mit meinem Kind über die Hausaufgaben streiten würde bzw. nicht so viel diskutieren müsste.\\
\end{enumerate}

\textbf{SNI:\textit{Subjective Norm - injunctive}}
\begin{enumerate}
    \item \textit{My friends and acquaintances would think it would be good if my child could use learning software to support homeschooling.}\\
        Meine Freunde und Bekannte fänden es gut, wenn mein Kind eine Lern-Software zur Unterstützung im Homeschooling nutzen würde.\\
\end{enumerate}

\textbf{SND:\textit{Subjective Norm - descriptive}}
\begin{enumerate}
    \item \textit{Most parents use (or would purchase) educational software to support their children with homeschooling at home.}\\
        Die meisten Eltern nutzen eine Lern-Software (oder würden sich diese anschaffen), damit ihre Kinder zu Hause beim Homeschooling unterstützt werden.
    \item \textit{Most parents think that educational software is a good way to support children in homeschooling.}\\
        Die meisten Eltern halten eine Lern-Software für eine gute Möglichkeit, um Kinder im Homeschooling zu unterstützen.\\
\end{enumerate}

\textbf{PBC:\textit{Perceived Behavioral Control}}
\begin{enumerate}
    \item \textit{I am sure that my child and I could use the learning software ("Flobi").}\\
        Ich bin mir sicher, dass mein Kind und ich die Lern-Software („Flobi“) nutzen könnten.
    \item \textit{I am confident that I could integrate the learning software ("Flobi") into our daily life at home.}\\
        Ich bin zuversichtlich, dass ich die Lern-Software („Flobi“) in unseren Alltag zu Hause integrieren könnte.
    \item \textit{It is up to me and my child to use the learning software (“Flobi”).}\\
        Es liegt an mir und meinem Kind, die Lern-Software(„Flobi“) zu nutzen.
    \item \textit{My child is in control of whether or not to use the learning software ("Flobi").}\\
        Mein Kind hat selbst in der Hand, ob es die Lern-Software („Flobi“) nutzt oder nicht.\\
\end{enumerate}

\textbf{BI:\textit{(Behavioral) Intention}}
\begin{enumerate}
    \item \textit{If I had the opportunity, I would use the learning software ("Flobi") to support my child in homeschooling.}\\
         Wenn ich die Möglichkeit hätte, würde ich die Lern-Software („Flobi“) zur Unterstützung meines Kindes im Homeschooling nutzen.
    \item \textit{I intend to use the learning software ("Flobi") to support my child in homeschooling as soon as it is available.}\\
        Ich habe vor, die Lern-Software („Flobi“) zur Unterstützung meines Kindes im Homeschooling zu nutzen, sobald diese verfügbar ist.\\
\end{enumerate}

\textbf{PB:\textit{Past Behavior}}
\begin{enumerate}
    \item \textit{I have already used learning software in the past few months (during Cororna time) to help my child in homeschooling.}\\
        Ich habe in den vergangenen Monaten (in der Cororna-Zeit) bereits Lern-Software genutzt, um mein Kind im Homeschooling zu unterstützen.\\
\end{enumerate}

\textbf{Expectations:}\\~
\textit{What would you expect from such software that supports your child in homeschooling?}\\
Welche Erwartungen hätten Sie an so eine Software, die Ihr Kind beim Homeschooling unterstützt?\\

\section{Semi-structured Interview}\label{app:children_survey}

\textbf{System Usability Scale - SUS:}
\begin{enumerate}
    \item \textit{Would you use Flobi on a regular basis?}\\
        Kannst du dir vorstellen, Flobi regelmäßig zu benutzen?
    \item \textit{Do you find it difficult to use Flobi?}\\
        Findest du es schwierig Flobi zu nutzen?
    \item \textit{Do you find it easy to use Flobi?}\\
        Findest du es leicht Flobi zu nutzen?
    \item \textit{Do you think you need help, e.g. from your parents, to use Flobi?}\\
        Denkst du, dass du Hilfe, z.B. von deinen Eltern, benötigst um mit Flobi umzugehen?
    \item \textit{Do you think that the things Flobi shows and says are well integrated in the program?}\\
        Findest du, dass die Sachen die Flobi zeigt und sagt gut zum gesamten Programm passen?
    \item \textit{Did you find parts of Flobi useless?  (If asked: So, that some things didn't fit together somehow?)}\\
        Hast du Teile von Flobi als sinnlos empfunden?  (Bei Nachfrage: Also, dass manche Sachen irgendwie nicht zueinander gepasst haben?)
    \item \textit{Do you think that most people learn quickly how Flobi works?}\\
        Denkst du, dass die meisten Menschen schnell lernen wie Flobi funktioniert?
    \item \textit{Do  you find the System of Flobi cumbersome?}\\
        Findest du die Bedienung von Flobi umständlich?
    \item \textit{Did you know what to do at all times?}\\
        Wusstest du zu jedem Zeitpunkt was zu tun ist?
    \item \textit{Did you feel like you had to learn a lot about the program before you could use Flobi?}\\
        Hattest du das Gefühl vorher viel über das Programm lernen zu müssen, bevor du Flobi nutzen konntest?\\
\end{enumerate}

\textbf{Questions regarding Flobi and the general system:}
\begin{enumerate}
    \item \textit{Did you like the point system?}\\
        Hat dir das Punktesystem gefallen?
    \item \textit{Did Flobi motivate you to stay focused on the tasks?}\\
        Hat dich Flobi motiviert, weiter konzentriert an den Aufgaben zu sitzen?
    \item \textit{Did you understand when Flobi praised you or gave you advice?}\\
        Hast du verstanden, wenn Flobi dich gelobt hat oder dir einen Rat gegeben hat?
    \item \textit{Did you think Flobi was fair?}\\
        Fandest du, dass Flobi fair war?\\
\end{enumerate}

\textbf{Open questions:}
\begin{enumerate}
    \item \textit{What would you change about Flobi if you could? (if no answer: ask specifically about design )}\\
        Was würdest du an Flobi ändern, wenn du könntest? (konkret über Design Nachfragen bei keiner Antwort)
    \item \textit{What would you change about the program if you could?}\\
        Was würdest du an dem Programm ändern, wenn du könntest?
    \item \textit{Gab es etwas, was du besonders gut fandest?}\\
        Gab es etwas, was du besonders gut fandest?
    \item \textit{Is there anything else you would like to share?}\\
        Gibt es sonst noch etwas, das du uns mitteilen möchtest?
\end{enumerate}

\printcredits

\bibliographystyle{cas-model2-names}

\bibliography{cas-refs}


\bio{}
Dr.-Ing Birte Richter: Birte Richter is an experienced researcher in the  Medical Assistive Systems Group at the Medical School OWL at Bielefeld University. She received her PhD in 2021 in the doctoral program Intelligent Systems.
From 2017 to 2020, she was a researcher in the Applied Computer Science Group at the Faculty of Technology at Bielefeld University, and from 2014 to 2017 part of the Cluster of Excellence Cognitive Interaction Technology (CITEC). Her research interests include human-agent interaction now, especially in the context of medical assistance systems, ranging from robots to smart home devices and apps, providing support for people in physical and cognitive tasks.
\endbio

\bio{}
M.Sc. Psych. Ira-Katharina Petras:
Since 2020, Clinical Psychologist and Research Associate at the University Clinic for Child and Adolescent Psychiatry and Psychotherapy at the Evangelisches Klinikum Bethel. 2017 to 2020 Research associate at the Department of Clinical Psychology at the University of Bremen. In training as a child and adolescent psychotherapist.
\endbio

\bio{} M.Sc. Ayla Luong: From 2020 to 2022, research assistant in the Department of Medicine OWL, Medical Assistance Systems. 2018 to 2020 researcher assistant in the Applied Computer Science Group at the Bielefeld University.
\endbio

\bio{}
Prof. Anna-Lisa Vollmer:
Anna-Lisa Vollmer is Professor for Interactive Robotics in Medicine and Care at the Medical School OWL at Bielefeld University, Germany. After receiving a PhD from Bielefeld University in cooperation with Honda Research Institute Europe, she was an Experienced Researcher with Angelo Cangelosi in the MSCA ITN RobotDoC, Robotics for the Development of Cognition, at Plymouth University, UK and was awarded a Starting Research Position working with Pierre-Yves Oudeyer at INRIA Bordeaux/ENSTA Paris, France. Her research interests include robot learning in interaction with lay users, social robotics, and the application of interactive robots in medicine and care.
\endbio

\bio{}
Prof. Dr. med. Michael Siniatchkin: Michael Siniatchkin is Director of the University Clinic of Child and Adolescent Psychiary and Psychotherapy, Protestant Hospital Bethel, University of Bielefeld and from the October 2023 Head of the Child and Adolescent Psychiatry clinic, RWTH Aachen, Germany. He studied medicine in Saratov, Moscow and Kiel and completed his residency in Child and Adolescent Psychiatry at Universities of Göttingen and Kiel, Germany. After being Deputy Director of Child Psychiatry in the Clinic in Frankfurt, he was a Head of the Institute of Medical Psychology, University of Kiel. His research focuses on neurobiological machanisms of child psychiatric disorders and development of innovative therapeutic options for children and adolescents including e-meantal health and application of technology (robotics) in child and adolescent psychiatry.  
\endbio

\bio{}
Prof. Britta Wrede:  Britta Wrede is head of the Medical Assistive Systems Group at the Medical School OWL at Bielefeld University. After receiving her M.A. and PhD from the Faculty of Linguistics in 1999 and the Technical Faculty in 2002 respectively, she investigated together with E. Shriberg dialog phenomena at the International Computer Science Institute (ICSI) in Berkeley. In 2019 she founded the new research group on Medical Assistive Systems at the Medical School OWL and, in 2022, spent a year at the University of Bremen to collaborate with Tanja Schultz and Michael Beetz on co-constructive task learning.  
In her work Britta Wrede has focused on understanding and modelling multimodal interaction strategies in human-agent interaction by applying and developing theoretical accounts on pragmatics in establishing mutual understanding especially in children. Britta Wrede is member of the DFG Senate committee for Graduate Schools. She is co-editor of the German Journal of Artificial Intelligence (Künstliche Intelligenz) and associate editor of the International Journal of Robotics Research (IJRR).
\endbio

\end{document}